\documentclass[apjl]{emulateapj}
\usepackage{natbib}
\usepackage{color}
\usepackage{hyperref}
\newcommand{\so}{3C~454.3}
\newcommand{\mg}{Mg~II$\lambda 2800$\AA}
\newcommand{\cnt}{UV-continuum~$\lambda 3000$\AA}
\newcommand{\mm}{$\lambda 8$mm}
\newcommand{\fe}{Fe~II}

\begin{document}

\title{Flare-like variability of the \mg\ emission line in the $\gamma$-ray blazar \so}

\author{J. Le\'on-Tavares\altaffilmark{1,2}, V. Chavushyan\altaffilmark{3}, V.  Pati\~no-\'Alvarez\altaffilmark{3}, E.  Valtaoja\altaffilmark{4}, T.  G.  Arshakian\altaffilmark{5,6}, L.~\v C.~Popovi\'c\altaffilmark{ 7,8,9}, M. Tornikoski\altaffilmark{2}, A. Lobanov\altaffilmark{10,11}, A. Carrami\~nana\altaffilmark{3}, L.~Carrasco\altaffilmark{3} , A.~L\"ahteenm\"aki\altaffilmark{2}}

\altaffiltext{1}{ Finnish Centre for  Astronomy with ESO (FINCA), University of Turku, V\"ais\"al\"antie 20, FI-21500  Piikki\"o, Finland \email{jonathan.leontavares@utu.fi}}
\altaffiltext{2}{ Aalto University Mets\"ahovi Radio Observatory,  Mets\"ahovintie 114, FIN-02540 Kylm\"al\"a, Finland}
\altaffiltext{3}{ Instituto Nacional de Astrof\'{\i}sica \'Optica y Electr\'onica (INAOE), Apartado Postal 51 y 216, 72000 Puebla, M\'exico}
\altaffiltext{4}{ Tuorla Observatory, Department  of Physics and Astronomy, University of Turku, 20100 Turku, Finland}
\altaffiltext{5}{ I. Physikalisches Institut, Universit\"at zu K\"oln, Z\"ulpicher Str. 77, 50937 K\"oln, Germany}
\altaffiltext{6}{ Byurakan Astrophysical Observatory, Byurakan 378433, Armenia and Isaac Newton Institute of Chile, Armenian Branch, Armenia }
\altaffiltext{7}{ Astronomical Observatory, Volgina 7, 11160 Belgrade 74, Serbia}
\altaffiltext{8}{ Isaac Newton Institute of Chile, Yugoslavia Branch, Belgrade, Serbia}
\altaffiltext{9}{Department of Astronomy, Faculty of Mathematics, University of Belgrade, Studentski Trg 16, 11000 Belgrade, Serbia}
\altaffiltext{10}{ Max-Planck-Institut f\"ur Radioastronomie, Auf dem H\"ugel 69, 53121 Bonn, Germany }
 \altaffiltext{11}{ Institut f\"ur Experimentalphysik, Universit\"at Hamburg, Luruper Chaussee 149, 22761 Hamburg, Germany}

\begin{abstract}
We report the detection of  a statistically significant  flare-like event in  the  \mg\  emission line  of \so\ during the  outburst of autumn 2010. The highest levels  of emission line flux  recorded over the monitoring  period (2008 - 2011)  coincide  with a  superluminal jet component traversing through the radio core. This finding crucially links the broad-emission line fluctuations  to  the non-thermal continuum emission   produced by relativistically moving  material in the  jet and hence to the presence of broad-line region clouds surrounding the radio core.  If the radio core were  located at several parsecs from the central black hole then our  results  would suggest the presence of  broad-line region material  outside the inner parsec where the canonical broad-line region is  envisaged to be located.  We briefly discuss the implications of   broad-emission line material ionized by non-thermal continuum   on the context of virial black hole mass estimates and gamma-ray production mechanisms.
\end{abstract}

\section{Introduction}

It has been long known that the optical emission of the flat spectrum radio quasar \so\  at $z=0.859$  is highly variable  \citep{angione_1968}, showing   structural and flux variability on its parsec scale jet \citep{pauliny_1987} and also correlated variability among different wave bands \citep{tornikoski_1994}.  The recurrent flaring  behaviour of \so\ over the last six years \citep[2005-2011,][]{fuhrmann_2006,giommi_2006,raiteri_2008a,raiteri_2008b,vercellone_2008,striani_2010,ackermann_2010,vercellone_2011}  has permitted the assembly of  an exquisite multiwavelength  time resolved  database \citep[e.g.][]{bonning_2009,raiteri_2011,sasada_2012} allowing to probe  intraday variability  in the source \citep[e.g][]{foschini_2010, gaur_2012}. Despite the precious  database compiled,  no general consensus about the location of the gamma-ray production zone in \so\  has been reached so far.  Some studies   \citep[e.g.][]{tavecchio_2010,poutanen_2010} favor the scenario where the gamma-ray emission in \so\ is generated close to the central black hole (BH) within the canonical broad-line region (BLR), which is   located within the inner parsec. However,   other works   \citep[e.g.][]{sikora_2008,jorstad_2010}  find that the scenario where gamma-rays are produced far from the central BH, within or downstream of the radio core at distances much larger than 1~pc,   best reproduces  the observations of \so.

In this work we  explore the variability of the  broad-emission lines in \so\  in order to use it as  an auxiliary piece of information to   probe  the geometry and physics of the innermost regions of \so\  and  to provide evidence for the above scenarios of the gamma-ray production.  Although \so\ has been monitored extensively at all wavelengths,  its  broad-emission lines have been scarcely systematically studied  \citep{raiteri_2008b,smith_2009,benitez_2010}. The present  work is the first  to address the variability of emission lines in \so\ during the $Fermi/LAT$ era with the  largest sample  of its optical spectra ever  compiled.

\section{Observations}

\begin{figure*}[t]
\center
\includegraphics[width=\textwidth]{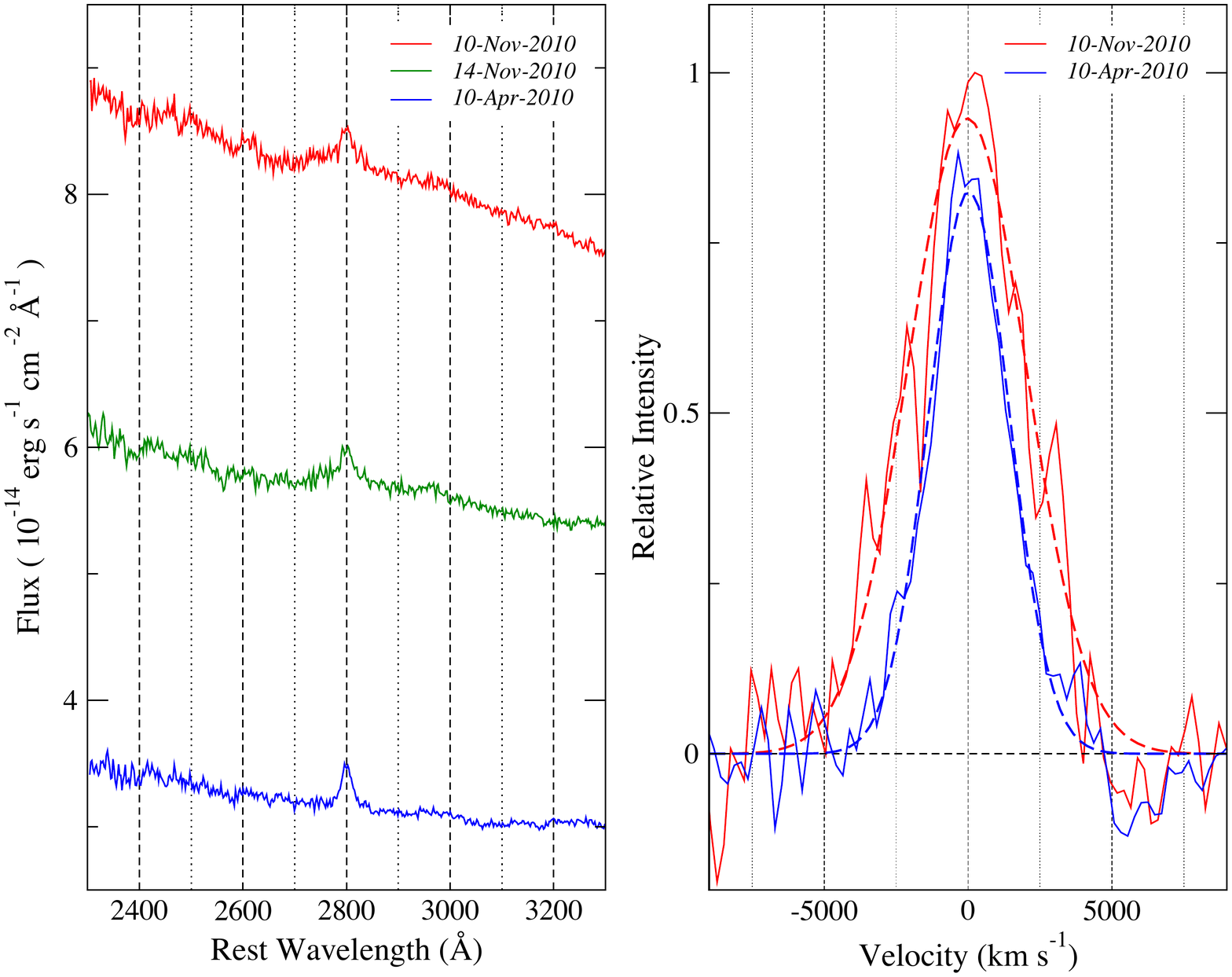}
\caption{ \emph{Left:} Rest frame  optical spectra of \so\  in three intensity states. As can be seen from the top spectrum, the \mg\  emission line is  detectable, despite the high levels of  optical continuum emission observed.   \emph{Right:} Comparison of the variations seen in the \mg\ profiles  (after continuum and Fe II subtraction) from April to November 2010. The observed spectra are shown in solid line  and a fitted gaussian to the profile are shown in dashed lines. Observing times are color coded as shown in the legend.  }\label{fig:spec}
\end{figure*}

\begin{figure*}
\center
\includegraphics[scale=0.49]{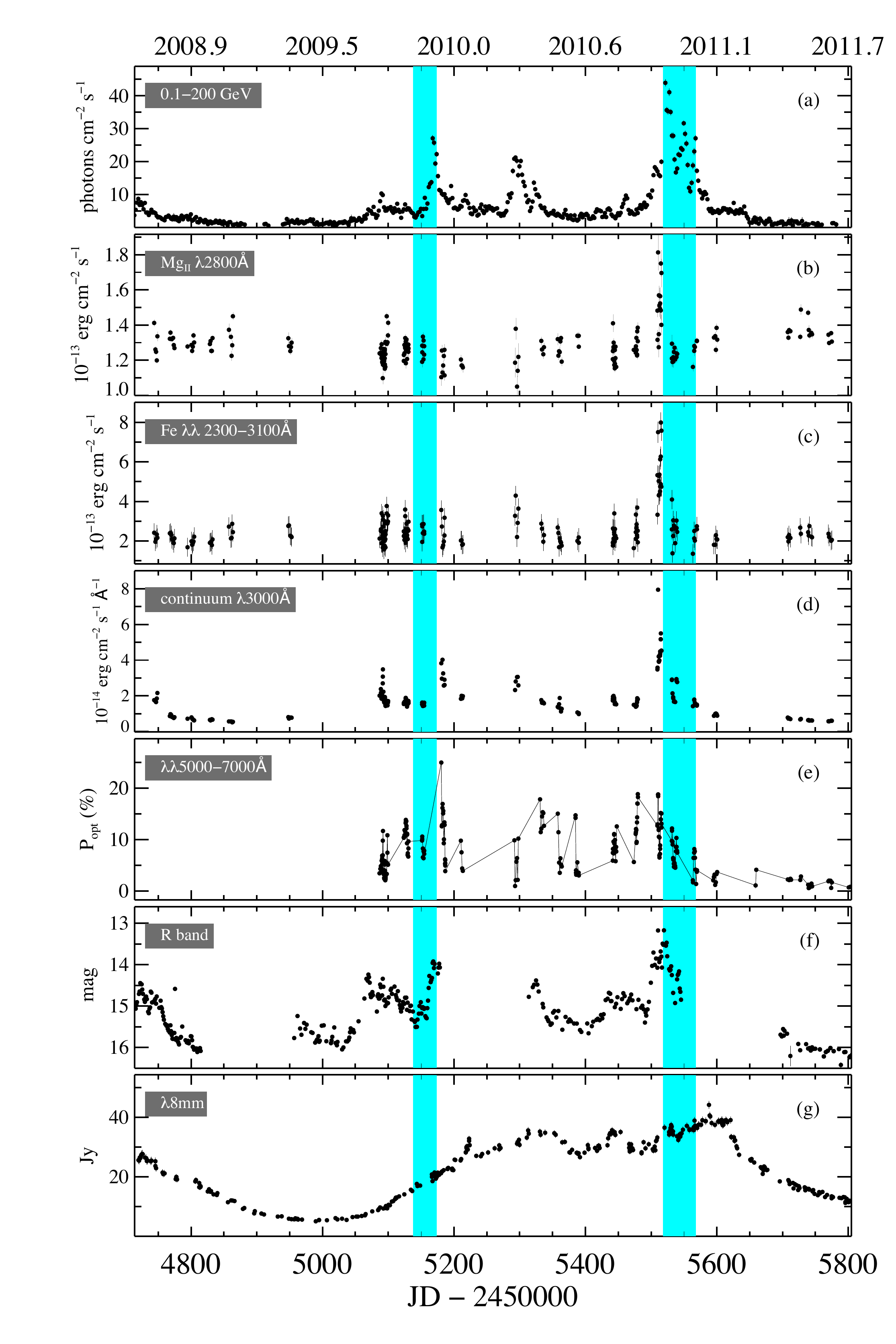}
\caption{Multiwavelength evolution of \so. The vertical stripes  show the time when new blobs were ejected from the radio core and their widths  represent the associated uncertainties. It should be noticed that the highest levels of \mg\ line flux occurred after   \mm\ flare onset, during an increase in the optical  continuum  and polarization percentage  and  before  the emergence of a new superluminal component from the radio core.}\label{fig:mw}
\end{figure*}

The optical spectra used in this work are taken from the Ground-based Observational Support of the Fermi Gamma-ray Space Telescope at the University of Arizona monitoring program\footnote{\href{http://james.as.arizona.edu/~psmith/Fermi/} {http://james.as.arizona.edu/?psmith/Fermi/Website}},  details on the observational setup and reduction process are presented in \citet{smith_2009}. In this work we only consider spectra that have been calibrated against the V-band magnitude. The spectra have been brought to the rest frame of the source,  a cosmological correction of the form $(1+z)^{3}$ has been performed and  no  correction for galactic reddening was applied. The left panel of Figure  \ref{fig:spec} displays three optical  spectra  of \so\  taken at three different periods of activity.  Here, we present  a brief description of our spectral fitting, which is based on least-squares minimization using the \texttt{MPFIT} package \citep{mpfit} and closely  follows the methodology described in \citet{torrealba_2012}. Firstly,  the featureless continuum  is approximated by a power law function and subtracted from the spectrum. Then,  the optical Fe~II emission is fitted  using the template of \citet{vestergaard_2001}.  After the local continuum and Fe II emission subtraction (see right panel of Figure \ref{fig:spec}), for all of the spectra, the \mg\ emission line flux was measured by integrating the line profile in the range $\lambda\lambda$2725-2875 \AA$\,$.  The spectra  in our database were taken with different slit sizes (different resolutions)  which prevents us to do an immediate analysis of the \mg\ FWHM evolution. Therefore, we  only measure the flux of Mg II line which is not affected by spectral resolution.

The errors associated  with the \mg\ flux measurements comprise three different error factors. The first error factor is the  random error due to dispersion of the spectra and the signal to noise ratio,   estimated as in   \citet{tresse_1999}. The second factor is  the error introduced by the subtraction of Fe emission, $\sigma_{Fe}= \frac{\sigma}{max_s}(F_{line}-F_{ironless})$, where $\sigma$ is the rms of the spectrum  after iron subtraction, $max_s$ is the line-peak in the continuum-less spectrum, $F_{line}$ is the line flux, and $F_{ironless}$ is the fraction of the line flux within  the range of integration where  no  iron subtraction was performed. The ratio $\sigma/max_s$ is basically the inverse of the S/N ratio of the line-peak, and the second term $F_{line}-F_{ironless}$ is introduced to scale the S/N  to the fraction of the line flux that could be affected by the iron subtraction.  The third error factor  is a result of  flux calibration, which is approximated by the RMS of flux measurements from  contiguous observing campaigns  (P. Smith, private communication). Here, we take a very conservative approach and approximate the flux calibration error  by the standard deviation of the flux over the three-year monitoring period  excluding the observing season showing  the highest levels of \mg\ line flux ($\geq 2\sigma$). The sum of the first and second error factors is considered here as the error of measurements ($\sigma_{F_{Mg}}$) and is presented in the data points shown in  Figures 1 to 4 and listed in Table~\ref{tbl-1}.  However,   the sum of all three sources of error, hereafter called total error ($\sigma_{F_{Mg}}^{total}$),  is used  in  our statistical analysis of variability.  The \cnt\ and \fe\ fluxes are also listed in Table~\ref{tbl-1}.

Figure \ref{fig:mw} shows the multiwavelength behaviour of \so\ during the \mg\ monitoring period. The gamma-ray light curve  from 0.1 to 200 GeV shown in panel $(a)$   was built  by  using data from the $Fermi$ Large Area Telescope (LAT) reduced and analyzed with the  \texttt{Fermi Science Tools v9r23p1} along with the latest diffuse model files. We  have used the user-contributed software to generate the spectral models and  produce the gamma-ray light curve. The \mg\ flux evolution is shown in panel $(b)$.  The light curve of the \fe\ emission is displayed in panel (c).  The \cnt\ flux shown in panel $(d)$ has been measured from the featureless continuum  after removal of Fe~II template, and the  error reported is  the RMS of the spectrum  within 2900-2950 \AA. The optical linear polarization data displayed in panel $(e)$ was also taken from the Steward monitoring program.  The $R$-band light curve shown in panel $(f)$ is taken from the Yale Fermi/SMARTS project\footnote{\href{http://www.astro.yale.edu/smarts/glast/tables/3C454.tab}{http://www.astro.yale.edu/smarts/glast/tables/3C454.tab}}.  The single dish monitoring data at 37 GHz (\mm ) presented in panel $(g)$  is adopted from the Mets\"ahovi quasar monitoring program  \citep{terasranta_1998}.   The vertical stripes in each panel  of Figure~\ref{fig:mw} show the ejection times of  new blobs from the \emph{radio core} \citep{jorstad_2012}. Hereafter, we refer to radio core as the unresolved feature in $\lambda7$mm (43~GHz) VLBI maps associated with the emergence of new superluminal components, unless the frequency of observation is explicitly stated.

\section{Variability of the  \mg\ emission line.}

\begin{figure}
\center
\includegraphics[width=\columnwidth]{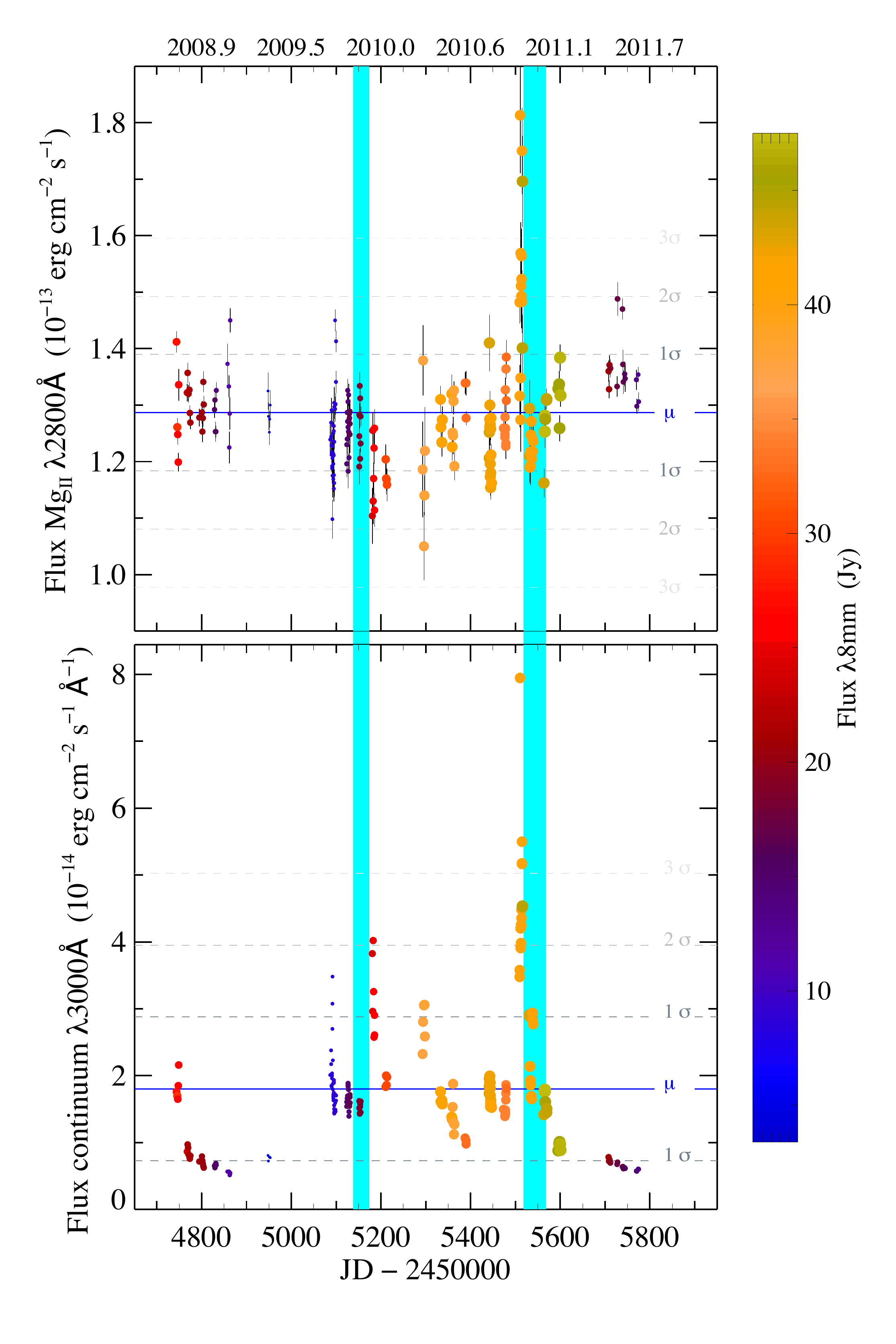}
\caption{ Flux evolution of  \mg\  emission line (\emph{top panel}) and \cnt\  emission (\emph{bottom panel}). For each panel, the solid (blue) horizontal line denotes the mean flux ($\mu$)  observed during  the monitoring period, whereas  dashed (gray)  horizontal lines show multiples of $\sigma$, where $\sigma$ is the standard deviation of the flux. In this work, we consider a significant flare if the levels of emission exceed 2$\sigma$. Symbol size and color are  coded according to the color bar displayed, where the larger and lighter the symbols,   the higher the  level of  \mm\ emission observed. The vertical stripes  are as in Figure~\ref{fig:mw}.  } \label{fig:evolution}
\end{figure}

\begin{figure*}
\includegraphics[width=\textwidth]{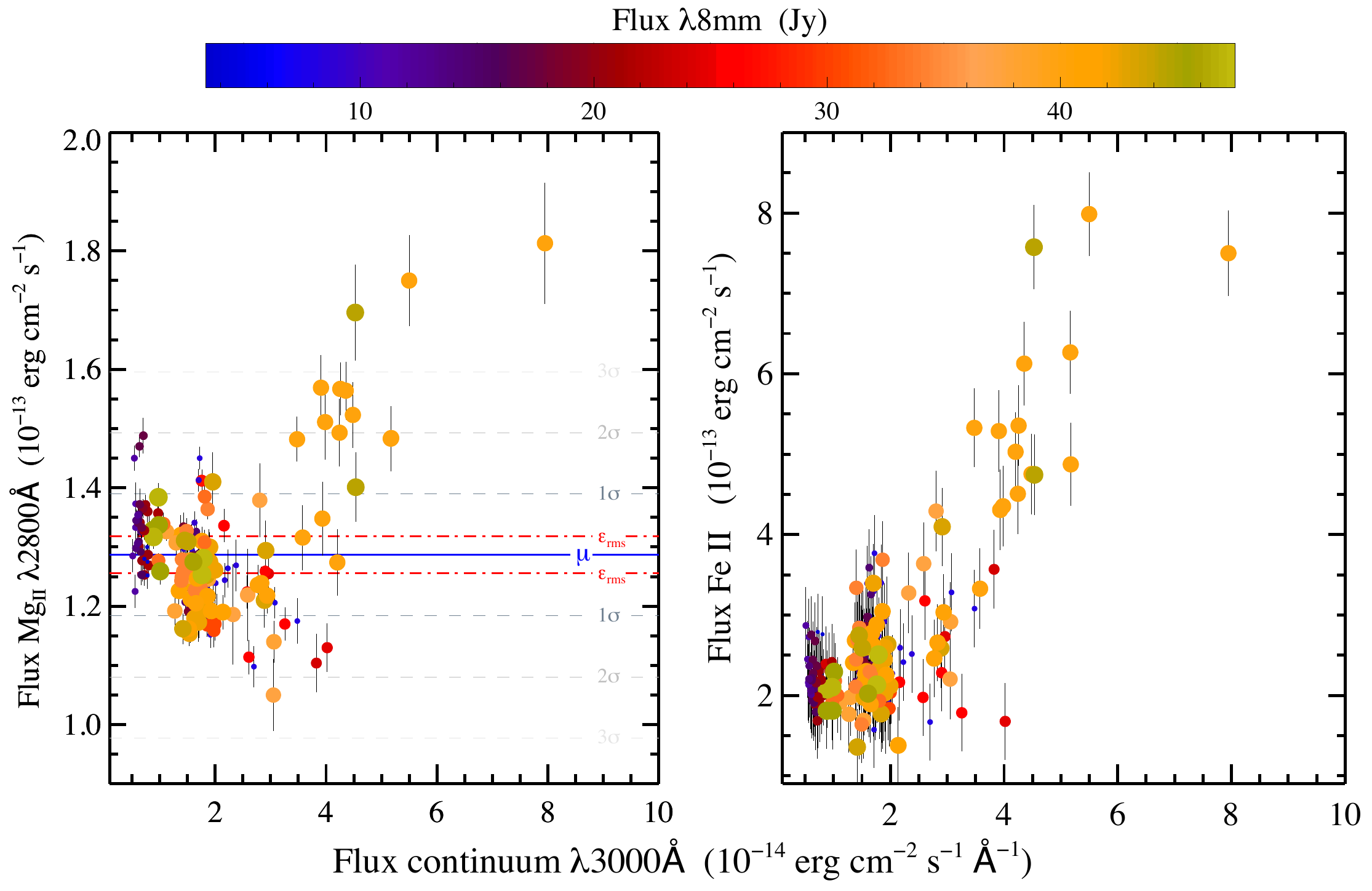}
\caption{The response of the \mg\ emission line (\emph{right}) and \fe\ emission (\emph{right}) to levels of \cnt\ continuum emission,  where the symbol color-size coding and nomenclature of horizontal lines is the same as in Figure \ref{fig:evolution}. The dot-dashed lines represent the  root mean square uncertainty ($\epsilon_{rms}$).}\label{fig:scatter}
\end{figure*}

The flux evolution of the integrated \mg\ emission line  (top panel)  and  the rest frame  \cnt\   (bottom panel) are shown in Figure \ref{fig:evolution}  and the data is displayed on Table \ref{tbl-1}. The significance of the fluctuations with respect to the mean flux ($\mu$)  are shown over-plotted as dashed lines. The  flux density  at 37~GHz  ($\lambda$8mm)  is  symbol size and color coded -- the larger and lighter the symbol, the more intense  the millimeter emission.   Variability of the  \mg\  emission can be already suggested from visual inspection of the light curve  and it is confirmed by  the large $\chi^{2}$ values obtained  with the  \texttt{MPFIT}  package   by fitting a constant flux. Under the assumption that the \mg\  light curve can be  represented by a constant line, the best fit shows a value  of   $\chi^{2}=291$. Hence, taking into account the degrees of freedom of the fit  (N$_{dof}$ =  205), the probabilities of getting this  $\chi^{2}$ value by chance (estimated with routine \texttt{MPCHITEST} of \texttt{MPFIT} library)  is $P < 1 \times 10^{-4}$. Such a small probability allows us to  reject the hypothesis that the emission from the \mg\  line did not vary over the last three years.

However,  no evidence of statistically significant variability in the H$\alpha $ flux of \so\ was found  by  \citet{raiteri_2008b}  after analyzing 16 NIR spectra taken over a six-month period in  2007 (12 June - 03 November) during a faint optical state. Similar  to  Figure 8 of  \citet{raiteri_2008b},  we show   the response of the  \mg\  line flux  to the  UV-continuum   $\lambda3000$\AA\    in   Figure \ref{fig:scatter}.  The root mean square uncertainty  $\epsilon_{rms}  = \sqrt{\Sigma^{N}_{i=1} ~ \frac{\epsilon^{2}_{i}} {N}}$ (where $\epsilon_{i}$ are the individual errors and $N$ is the number of spectra) of the \mg\ line flux is over-plotted  as a dot-dashed line. As can be seen, the root mean square of the uncertainties in the measurements of the \mg\ line flux is considerably smaller than its standard deviation  ($\sigma  > \epsilon_{rms}$). Following  the same  reasoning that \citet{raiteri_2008b} used to discard  variability in the H$\alpha$ line emission of \so, we then find that  over the period of time considered in this work (2008-2011) the \mg\ line flux variability is  well resolved  and  therefore statistically significant.

We find that during the monitoring period,  the spectral variation   ($F_{max} / F_{min}$)  of the  \mg\ line   is 1.7.   The peak-to-peak change in the \mg\ light curve is  60\% of the mean line flux, while the minimum  and maximum \mg\ line flux reached   are  about  20\% and 40\%  of the mean line flux, respectively.      Because  of  the sparsely sampled \mg\ light curve we are only able to  estimate an upper-limit to its variability timescale: $\tau \leq 1$ month.

Now that we have shown that variability of \mg\ in \so\ is statistically significant, it is instructive to investigate  whether  there is a  causal link between major  \mg\  line flux  variations and variability at other wavelengths, see Figure \ref{fig:mw}.  We have also intentionally added a color bar in Fig \ref{fig:evolution} and \ref{fig:scatter} to show that the highest levels of \mg\ line flux ($\geq 2\sigma$), occurred over a short period of time ($5500 < JD - 2450000 < 5600$),  coincide with the   \mm\ fluxes above 35 Jy  (light red to yellow in color scale).  The vertical stripes in Figure \ref{fig:mw} and \ref{fig:evolution} show the times when new blobs were ejected from the radio core, thus suggesting that the highest levels of \mg\ line flux occurred  after the onset of a mm flare and  before a new component was seen for the first time leaving the radio core.   A mm flare presumably starts to rise as the moving disturbance enters the radio core \citep{leontavares_2011}  and by definition,  the ejection time indicates  when such component passed through the middle of the radio core; see Figure 3 in \citet{leontavares_2012}. Therefore, the observed  causality of events   suggests that the highest levels of \mg\ emission line flux were reached when a superluminal jet component was traversing through the radio core

 How can we be sure that the ionizing continuum was produced within the radio core and not  by other thermal sources like the  accretion disk?   The high degree of linear polarization (see panel $e$ in Figure~\ref{fig:mw}) together with the  short time scale of variability and  the presence of a gamma-ray counterpart,  suggest that the major  \cnt\ flare  had a non-thermal \textbf{origin}.    This can be taken as  an evidence  for a non-thermal  source  being  responsible for the  ionization of the BLR clouds.  In  further support of this suggestion, in the left panel of  Figure~\ref{fig:scatter}  we can identify  a correlation between \mg\ and \cnt\  ($\rho\sim0.4$) which   becomes significant ($P  < 0.05$)   when the \mm \ emission rises above 30 Jy. In addition,  we also find a significant correlation between  \fe\  and   \cnt\  emission (see  right panel in Figure~\ref{fig:scatter})  in consistency with \citet{benitez_2010}. The latter  becomes tighter  with increasing \mm\ emission, thus also supporting the scenario in which  a complex structure of  the BLR is ionized by the jet.   In order  to  determine the  kinematical properties of the BLR clouds  illuminated by the inner jet, we will perform a  detailed analysis of the emission line profiles (studying the FHWM \mg\ evolution)  in a forthcoming paper.

At this point, a discerning reader might  ask,  why was not  a \mg\  flare associated to the  component ejected in late 2009? Although   explaining the physics of correlated multiwavelength behaviour in \so\ is out of the scope of this work  \citep[and we refer the reader to other recent  studies, e.g. ][]{bonnoli_2011,raiteri_2011,wehrle_2012},   we speculate that the detection of  high levels of \mg\  emission might be related to the amount of energy dissipated  by the jet component when passing through the radio core.  From Figure \ref{fig:mw}, it can be gleaned that the energy dissipation in the events of 2009  and 2010 took place downstream and upstream  of the radio core, respectively.   By using the maxima of  \cnt\ emission as a proxy for the energy dissipated, we  see that the event of 2010 released twice as much energy than the event from 2009.  If the amount of energy released at the premises of the radio core depends on whether the energy dissipation occurs upstream or downstream of the radio core   is something that deserves further investigation.


\section{Summary and Discussion}

The  optical spectra of \so\  have been acquired,  as part of  the  Ground-based Observational Support of the Fermi Gamma-ray Space Telescope at the University of Arizona monitoring program,  over a period of three years (2008-2011) and due to its  redshift ($z=0.859$)  we have had access to the middle-UV region of the spectrum, allowing us to  monitor its \mg\ broad-emission line and  adjacent \cnt. We summarize our results as follows:

\begin{itemize}

 \item We find  a statistically significant  flare-like event  in the   \mg\ light curve of \so. The  maximum \mg\ line flux recorded is above  40\% of the mean flux and  the spectral variation ($F_{max} / F_{min}$)  of the  \mg\ line   is 1.7. However, due to the sparsely sampled \mg\ light curve we are only able to  estimate an upper-limit to its variability timescale $\tau \leq 1$ month.

\item The highest levels of \mg\ line flux ($\geq 2\sigma$) occurred after   \mm\ flare onset, during an increase in the optical polarization percentage,  before  the emergence of a new superluminal component from the radio core  and within the largest $\gamma$-ray flare ever seen.   This finding crucially links the broad-emission line fluctuations  to  the non-thermal continuum emission   produced by relativistically moving  material in the  jet and hence to the presence of broad-line region clouds surrounding the radio core.

\end{itemize}

The results presented  above indicate that the  radio core   plays  a pivotal role  in  the energy release of \so. However, there is no general consensus about the true nature -- and location --  of the radio core in radio-loud AGN.  Some studies propose  that the radio core  is  a recollimation shock  and a genuine stationary feature along the parsec scale jet \citep[e.g.][]{gomez_1995} thought to be located at a considerable distance from the black hole. On the other hand, recent observational evidence  suggests  that  for  sources with a  misaligned jet (i.e. M~87)  the radio core might coincide with the BH \citep{hada_2011}.  Recenlty, \citet{pushkarev_2012} studied the nuclear  opacity in the parsec scale jet of a large sample of AGN. The authors report  that, for \so, the  distance from the radio core at 15~GHz to the apex of the jet  is  about  20 parsecs. Although  the radio core at 43~GHz should be located closer to the BH,  the findings of    \citet{pushkarev_2012}  encourage the view that the radio core  in \so\ is located well downstream  (outside) the canonical BLR.

Despite the true nature and location of the radio core in \so\, the most important implication of our main finding  --  jet  powering  broad-emission lines during  intense outbursts --  is the presence of broad-line region material surrounding the radio core. If the radio core is embedded within the canonical BLR, then the ionization of the BLR by the jet would make the mirror model  to the production of gamma-rays as proposed by \citet{ghisellini_1996} feasible.  Moreover, a likely interaction of BLR clouds with the base of the jet might contribute to the observed levels of high-energy emission \citep{araudo_2010, boschramon_2012}.  Conversely, if the radio core is located far from the  BH, then broad-line region material should be present at distances of parsecs from the BH.  

Observational evidence for the presence of  BLR material  located at parsec scales down the radio core has been found  by   coordinated spectroscopic and VLBI monitoring studies.  More specifically, \citet{arshakian_2010} and \citet{leontavares_2010} found that for the radio galaxies 3C~390.3 \footnote{ \url{http://www.metsahovi.fi/~leon/movies/3c3903.gif}} and 3C~120 \footnote{\url{http://www.metsahovi.fi/~leon/movies/3c120.gif}} the  variable optical continuum starts to rise  when a new superluminal component leaves the radio core seen at 15~GHz  and  its maximum occurs when the  component  passes through a stationary feature located  downstream of  the radio core. Since these two radio galaxies are known to reverberate \citep{shapovalova_2010,grier_2012}, in the sense that the H$\beta$ broad-emission line responds to changes in the optical continuum, the authors conclude that the jet can power a significant amount of broad-line emission particularly during strong continuum flares in these objects.

The above  results suggest the presence of an additional component of the BLR,  dubbed as \emph{outflowing BLR},  which in effect might be filled with BLR material dragged by the relativistic jet as it propagates downstream of the BH or perhaps  could be a sub-relativistic outflow arising from an accretion-disk  wind.  The notion of a dynamic  and extended BLR has been previously proposed in several works \citep[e.g.][]{popovic_2001,elitzur_2006}.  Although these studies have focused on non-blazar sources,  the presence of an outflowing BLR  has already been suggested  for  sources  like 3C~273 \citep{paltani_2003} and \so\ \citep{finke_2010}    where it  could serve  as an  alternative source of seed photons  for the Inverse Compton scattering  as proposed by \citet{leontavares_2011}.

Regardless  of whether the radio core is close (1 pc $<$ ) or far ($\gg$1 pc) from the BH,  the fact that broad-emission lines respond to changes of the non-thermal continuum prevents us to use the single epoch virial BH mass estimates in this source because the latter assumes:  (i) a single localized ionization source (i.e.  accretion disk) and (ii)   virial equilibrium of the BLR clouds.  The latter assumptions cannot be fulfilled during episodes of strong flaring activity  in \so, hence the  ionization of BLR clouds by non-thermal emission  might  introduce uncertainties to the BH mass estimates derived by  assuming  virial equilibrium of the BLR. Alternative scaling relations to weigh the BH in strongly beamed sources  are discussed  and implemented in \citet{leontavares_2011_mbh}.

\acknowledgments{We thank the anonymous referee for  her/his positive and helpful  comments.  We are thankful to  P. Smith for his help on the spectral analysis. This work was supported by CONACyT research grant 151494 (M\'exico) V. PA. acknowledges support from the CONACyT program for PhD studies.  Data from the Steward Observatory spectropolarimetric monitoring project were used. This program is supported by Fermi Guest Investigator grants NNX08AW56G, NNX09AU10G, and NNX12AO93G. This work was partially supported by the Deutsche Forschungsgemeinschaft (DFG), project number Os 177/2-1. The Mets\"ahovi team acknowledges the support from the Academy of Finland
(project numbers 212656, 210338, 121148).  L.C.P. is supported by Ministry of Science and Education of R. Serbia, through project 176001  "Astrophysical spectroscopy of extragalactic objects'.  }

\clearpage

\begin{deluxetable}{ccrrrrrrrrcrl}
\tabletypesize{\scriptsize}
\tablecaption{Measurements\label{tbl-1}}
\tablewidth{0pt}
\tablehead{
\colhead{JD-2450000} & \multicolumn{2}{c}{\mg} & \multicolumn{2}{c}{\fe} & \multicolumn{2}{c}{\cnt} \\
                                         &   flux (10$^{-13}$ erg cm$^{-2}$ s$^{-1}$) & error                      &  flux (10$^{-13}$ erg cm$^{-2}$ s$^{-1}$ )&error                       & flux(10$^{-14}$ erg cm$^{-2}$ s$^{-1}$ \AA$^{-1}$ )&error }

\startdata
4743.8350&   1.412&  0.019&  2.418&  0.478 & 1.759 & 0.005\\
4745.7251&   1.261&  0.016 & 1.962 & 0.476  &1.698  &0.005\\
4746.7104 & 1.248 & 0.018 & 2.094 & 0.477&  1.648 & 0.005\\
4747.8491 &1.199 & 0.016 & 2.333 & 0.474  &1.850 & 0.005\\
4748.7373  &1.336  &0.028  &2.164 & 0.479  &2.159  &0.009\\
4767.6724   &1.322  &0.016 & 2.389 & 0.473  &0.864 & 0.005\\
4768.6597   &1.357 & 0.018  &2.416 & 0.473 & 0.971 & 0.006\\
4769.8086   &1.321 & 0.014 & 2.308 & 0.474 & 0.923 & 0.004\\
4770.7051   &1.320 & 0.016 & 2.101 & 0.475 & 0.821 & 0.004\\
4772.6831   &1.327 & 0.014 & 2.020 & 0.474  &0.788 & 0.004\\
 \enddata

\tablecomments{Table \ref{tbl-1} is published in its entirely in the 
electronic edition of the {\it Astrophysical Journal Letters}.  A portion is 
shown here for guidance regarding its form and content.}

\end{deluxetable}

\newpage

\bibliography{ms}

\end{document}